# Can Personalized Medicine Coexist with Health Equity? Examining the Cost Barrier and Ethical Implications

Kishi Kobe Yee Francisco[1†], Andrane Estelle Carnicer Apuhin[1†], Myles Joshua Toledo Tan[1,2,3,4,5,6,7], Mickael Cavanaugh Byers[8], Nicholle Mae Amor Tan Maravilla[7], Hezerul Abdul Karim[9*], Nouar AlDahoul[10]

[1] Biology Program, College of Arts and Science, University of St. La Salle, Bacolod City, Negros Occidental, Philippines
[2] Department of Electrical and Computer Engineering, Herbert Wertheim College of Engineering, University of Florida, Gainesville, Florida, United States of America
[3] Department of Epidemiology, College of Public Health & Health Professions and College of Medicine, University of Florida, Gainesville, Florida, United States of America
[4] Department of Natural Sciences, College of Arts and Science, University of St. La Salle, Bacolod City, Negros Occidental, Philippines
[5] Department of Chemical Engineering, College of Engineering and Technology, University of St. La Salle, Bacolod City, Negros Occidental, Philippines
[6] Department of Electronics Engineering, College of Engineering and Technology, University of St. La Salle, Bacolod City, Negros Occidental, Philippines
[7] Yo-Vivo Corporation, Bacolod City, Negros Occidental, Philippines
[8] Department of Civil and Coastal Engineering, Herbert Wertheim College of Engineering, University of Florida, Gainesville, Florida, United States of America
[9] Faculty of Engineering, Multimedia University, Cyberjaya, Selangor, Malaysia
[10] Department of Computer Science, Division of Science, New York University Abu Dhabi, Saadiyat Marina District, Abu Dhabi, United Arab Emirates
[†] These authors have contributed equally to this work and share first authorship.

*Correspondence:
Hezerul Abdul Karim [hezerul@mmu.edu.my](hezerul@mmu.edu.my)

## 1. Introduction: The Promise and Challenge of Personalized Medicine

The rapid evolution of biomedical data and computational tools is driving transformative changes in healthcare, with personalized medicine (PM) at the forefront (Tawfik et al., 2023). Often called precision medicine, PM leverages an individual's unique genetic, environmental, and lifestyle data to guide treatment decisions (Goetz & Schork, 2018), moving beyond the one-size-fits-all approach to offer tailored medical interventions. This strategy has generated widespread interest due to its potential to improve diagnostic accuracy, speed up disease prevention, and deliver effective treatments that minimize the need for trial and error. Proponents argue that PM could reduce long-term healthcare costs by focusing resources on specific, targeted therapies (Lu et al., 2023). However, implementing this high-tech approach requires significant financial and technological investments, which raises important questions about its accessibility and potential impact on health equity.

The promise of PM to revolutionize healthcare is currently most evident in high-income countries (HICs) where funding, infrastructure, and access to cutting-edge technology are more readily available (Alyass et al., 2015). For instance, genomics and artificial intelligence (AI)-enhanced diagnostics have enabled these countries to implement PM initiatives that improve outcomes in areas like oncology and rare genetic disorders (Khoury et al., 2022). Yet, low- and middle-income countries (LMICs), which face systemic healthcare barriers, are largely excluded from these advancements. In many LMICs, healthcare systems are under-resourced, with limited access to basic services and a shortage of medical professionals. These constraints make it challenging to adopt the resource-intensive tools and infrastructure required for PM, potentially creating an uneven global landscape where only affluent populations benefit from these advances (Trabelsi et al., 2024).

Central to this concern is the cost associated with PM. Precision treatments, particularly those involving genetic profiling, biomarker testing, and AI-driven diagnostics, entail substantial expenses for both development and implementation (Mlika et al., 2015). The process of drug discovery alone, with its reliance on genomic research and targeted therapies, can be cost-prohibitive, often reaching billions of dollars for a single therapeutic solution. Advanced gene therapies and specialized treatments, such as those in oncology, illustrate this financial burden, as they frequently require not only expensive drugs but also sophisticated diagnostic tools and highly trained personnel (Lu et al., 2023; Wong et al., 2023; Masucci et al., 2024). As these high costs trickle down to patients, they can become an insurmountable barrier, particularly in resource-poor settings where even basic healthcare remains a luxury.

Beyond financial constraints, PM poses ethical questions about fairness and justice. Ethical theories, such as Rawls' principle of fairness, Kantian ethics, and utilitarianism, provide frameworks to consider these dilemmas. Rawls' "difference principle" suggests that societal inequalities are only justifiable if they benefit the least advantaged ( Drane 1990; Daniels, 2008; Marseille & Khan, 2019); applied to PM, this principle challenges the notion of deploying advanced medical treatments that primarily benefit those who can afford them, potentially worsening health inequities. Kantian ethics, with its emphasis on universal human dignity, implies a moral obligation for healthcare systems to ensure that PM is accessible to all individuals, respecting the inherent worth of every person. Similarly, utilitarianism argues for maximizing overall well-being and could suggest that focusing on high-impact, cost-effective treatments might better serve global populations. Together, these theories underscore the importance of implementing PM in ways that enhance rather than undermine health equity.

This paper seeks to answer a critical question: Can the promising future of personalized medicine (PM) coexist with a commitment to health equity? Addressing this requires a comprehensive examination of PM's cost barriers, the infrastructure needed for equitable access, and the socio-political factors influencing healthcare delivery worldwide. To explore the implications of PM, this paper draws on insights using a transdisciplinary approach. It examines perspectives from disciplines such as genetics, global health, ethics, economics, and health policy to gain a comprehensive view. Unlike a traditional interdisciplinary approach, analyzing insights from these various fields through a transdisciplinary lens allows for a holistic understanding of PM's potential and the existing challenges beyond simply making unified connections (Choi and Pak, 2006). This approach acknowledges the diversity of viewpoints among different populations and stakeholders, ensuring an understanding of how to adopt PM efficiently and effectively while also addressing existing health inequities.

Through a review of ethical considerations and case studies from both high-income countries (HICs) and low- and middle-income countries (LMICs), this paper explores how innovative approaches could help bridge the gap between technological advancement and equitable healthcare access. Global health initiatives, supported by policies that emphasize collaboration and shared resources, have the potential to make PM a viable and inclusive approach to healthcare. In addressing these questions, this paper aims to propose strategies that ensure PM does not widen existing health disparities but instead becomes a tool for universal health improvement. By prioritizing infrastructure development, training, and financial investment in LMICs, PM can align with principles of health equity, creating a future where healthcare innovation benefits everyone, regardless of socio-economic status.

## 2. The High Cost of Personalized Medicine

Many have envisioned the positive transformation of healthcare through major advancements in science and technology, such as advancements in DNA sequencing tools, AI-assisted diagnostics, and precision treatments. Yet, this vision fails to account for the significant value-added costs required to materialize it. As Lu et al. (2023) point out, while scientific progress continues to accelerate, it ironically exacerbates core issues like affordability and access disparities that further challenge the delivery of health equity.

The same is true for personalized medicine; while it boasts advantages in its implementation and integration within clinical practices, one of the major obstacles to this advancement is primarily economic in nature (Jakka & Rossbach, 2013). With many factors contributing to the higher costs associated with accessing its declared benefits, Vellekoop et al. (2022) state that PM is significantly found to contribute to higher costs in healthcare and treatment, even though it also improves medical outcomes. Several key factors lead to increased financial burdens for personalized treatments, including the high costs of developing drugs and other therapeutics.

The drug development process involves different stages, each incurring substantial expenses, from preclinical research and drug development to clinical trials. This includes additional costs associated with collaborations between health systems and industry (Lu et al., 2023). Over the last few decades, the costs of developing drugs have risen significantly. As research and development efforts have expanded, the prices of these treatments have also begun to skyrocket (Sertkaya et al., 2024). Developing new and effective drugs is crucial for tailoring treatments to an individual's unique genetic profile, which is essential in personalized medicine. However, the exorbitant costs required to create a new drug increase further when the costs of failure are included (Rajkumar, 2020), with evidence suggesting that it averages around $314 million USD to $2.8 billion USD (Wouters et al., 2022).

Oncological treatments are a prominent example of the high-cost application of personalized medicine. The costs associated with cancer treatments, such as anticancer drugs and gene therapy, continue to increase (Leighl et al., 2021). Although the costs of these anticancer drugs vary globally, in the United States alone, their prices can reach up to $100,000 USD per year, with many requiring an adjuvant drug for increased efficacy, further driving up cancer treatment expenses (Workman et

al., 2017). Moreover, cancer drugs are initially introduced in the U.S., establishing high standard market prices that drive up costs worldwide (Leighl et al., 2021).

Another major contributor to the high costs of personalized treatments is the expense of advanced technologies required for its individualized approach. Personalized medicine relies heavily on technologies for exploring biomarkers, genetic, and molecular profiling. Meckley & Neumann (2009) emphasize that PM strategies are largely driven by these technologies, which can only be feasible with significant financial investments, leading stakeholders to increase personalized therapy prices to achieve substantial returns (Masucci et al., 2024). Gene therapies targeting specific diseases are also known for their high costs (Wong et al., 2023).

One example of this is the varying market prices for trastuzumab (Herceptin) for HER2-positive breast cancer, with 440 mg costing up to $1,852.07 USD in Taiwan alone (Diaby et al., 2020). Similarly, the price of trastuzumab deruxtecan (Enhertu), based on the Canadian Agency for Drugs and Technologies in Health (2023), is around $2,440.00 CAD per 100 mg, with an estimated annual cost of $165,949 CAD per treatment. In 2015, the UK National Health and Care Excellence ultimately stopped funding trastuzumab due to its high expense, as its retail price amounted to £90,000 GBP per patient (Kristin et al., 2022). While studies suggest that these cost burdens are evident in high-income countries (Stefanicka-Wojtas & Kurpas, 2023; Masucci et al., 2024), they are felt even more acutely in low- and middle-income countries (LMICs). For example, in the Philippines, the cost of sourcing this drug alone is enough to drain the entire national budget allocated for other drug acquisitions (Genuinino et al., 2019).

As Peters et al. (2008) demonstrate, poverty-stricken developing countries already suffer from underfunded healthcare systems and limited access to health services. Furthermore, the lack of advanced genetic testing facilities, adequate funding, research-focused education, health literacy, and a skilled workforce continues to hinder LMICs in overcoming persistent barriers to health equity (Alyass et al., 2015; Drake et al., 2018). For instance, there is a clear gap in genomic laboratory establishment in LMICs, with countries like Tunisia often relying on external funding, international collaborations, and the need to send samples abroad for high-throughput sequencing (Trabelsi et al., 2024). In the Philippines, genome sequencing services are limited to only a few national laboratories (Philippine Council for Health Research and Development, 2022; Philippine Genome Center Mindanao, 2023), and these services come at significantly high costs. Existing geographical barriers also make it challenging to access these facilities (Taruscio et al., 2023). Furthermore, most developing nations face clear inaccessibility to new, life-saving drugs. Challenges such as unestablished drug regulatory programs, logistics, weak health systems, and resource scarcity make acquiring these medications nearly impossible (Adeniji et al., 2021; Yenet et al., 2023).

The need to establish infrastructure and train skilled workers to build capacity for personalized medicine in LMICs is another pressing challenge contributing to high costs, requiring substantial financial investments and collaboration (Taruscio et al., 2023; Oduoye et al., 2024). Despite high-income countries starting to benefit from personalized medicine through genetic insights, the reduced costs for next-generation sequencing machines, and support from global organizations like the World Health Organization, many LMICs still cannot afford the necessary infrastructure to implement these innovations effectively (Deminco et al., 2022).

To date, numerous studies detail how LMICs will bear the brunt of the additional financial burden of personalized medicine (Agyeman et al., 2015; Alyass et al., 2015; Chong et al., 2018; Wang et al., 2023; Tawfik et al., 2023; Sarwar et al., 2023). While high-income countries struggle to manage the exorbitant costs that come with this evolving healthcare landscape, LMICs are still grappling with existing barriers to accessing adequate treatments to promote health, only to face greater challenges as personalized medicine moves farther out of reach.

In this context, these challenges reveal the broader implications personalized medicine could have on health equity. As countries worldwide confront existing disparities, compounded by the risks personalized medicine may impose, it becomes increasingly clear that health equity must be defined and examined to understand its intersection with the evolving healthcare landscape. This underscores how health equity plays a pivotal role in ensuring that personalized medicine is accessible to all, addressing gaps in access, treatments, and outcomes across diverse populations.

## 3. Health Equity: Definitions and Challenges

The pursuit of global health equity remains the central focus of global health initiatives. While health is a significant and fundamental right, the challenges to achieving it are still profoundly complex, especially in many developing countries. Definitions of health equity may vary, but with fairness and justice as its main theme, Braveman et al. (2018) define it as a principle where all people have equal chances of attaining their best possible level of health, free from barriers that could prevent this. Health equity essentially strives to remove impediments to accessing adequate and efficient healthcare, which hampers individuals' chances of achieving full health potential (World Health Organization, 2010). However, due to its transdisciplinary nature, it is prone to various challenges. Several studies (Peters et al., 2008; Braveman et al., 2011; Mills, 2014; Biglan et al., 2023) have shown that issues like resource allocation, policies, and agreements among global communities, as well as systemic factors such as discrimination based on gender, race, and socioeconomic status, hinder health equity, especially in LMICs.

In recent years, the COVID-19 pandemic has further exposed stark disparities in healthcare access worldwide, particularly between high-income countries and LMICs. Challenges in healthcare accessibility, vaccine procurement, treatments, and lack of essential resources highlight that the world is still far from achieving health equity (Jensen et al., 2021). Inequities in vaccine distribution are a clear example; Bayati et al. (2022) described that vaccine administration was 69% higher in developed countries than in developing countries. Despite global health initiatives by the United Nations to distribute vaccines more widely in low- and middle-income countries, these efforts were still found to be insufficient (Tatar et al., 2022). Additionally, the delayed access to vaccines by LMICs resulted in worsened health outcomes, increased infection rates, and higher mortality than in wealthier nations (Duroseau et al., 2022). These disparities underscore how economic status impacts essential health outcomes; even after vaccines became available in LMICs, there remains a significant need to ensure that all countries, regardless of wealth, have fair and easy access to health interventions.

Today, increasing barriers to health equity reveal the limitations of global health initiatives. These disparities have been exacerbated by the challenges of the pandemic, recent economic instability, and escalating political conflicts, leaving LMICs and low-income populations in wealthier countries grappling with food insecurity, housing instability, and limited access to hygiene facilities (Shadmi et al., 2020; Khorram-Manesh et al., 2023; Shorrab et al., 2024). Moreover, the global shortage of healthcare workers has only intensified these burdens. The World Health Organization projected in 2014 that by 2030, 18 million additional health workers would be needed in resource-limited countries to address these shortages (World Health Organization, 2016). Meanwhile, 17% of global healthcare workers are expected to retire in the next 10 years (De Vries et al., 2023).

Younger generations are increasingly deterred from entering the health profession due to factors such as low job status, poor compensation, and better career prospects elsewhere, not to mention the financial burden of nursing training (Grant-Smith & de Zwaan, 2019; Morley et al., 2024), rising living costs (Alibudbud et al., 2023), and the high costs of medical school (Burr et al., 2023). Consequently, numerous studies report that healthcare professionals in LMICs are often motivated to migrate to other countries for better job opportunities, greater job security, and higher salaries compared to their home countries (Ikhurionan et al., 2022; Toyin-Thomas et al., 2023). This migration helps address the shortage of health professionals in many high-income countries (Eaton et al., 2023) but poses significant threats to the healthcare systems they leave behind. This trend constitutes a substantial loss to already strained public health infrastructures, potentially widening health equity gaps.

Without an adequate workforce, vulnerable populations face even greater risks of inadequate healthcare. Although some argue that LMICs benefit economically from remittances sent by healthcare workers abroad (Gomes et al., 2024), this does not offset the shortage itself. For instance, this shortage limits the capacity to provide essential health services. Haddad et al. (2023) explain that nurses are vital to healthcare service delivery, and their absence or shortage results in worsened health outcomes, including increased disease burden and mortality risk. Similarly, Mullan (2005) highlights that the shortage of physicians in low-income countries is a pressing issue, reducing capacity in health emergency response and hindering disease prevention programs.

As Jensen et al. (2021) further discuss, the ability to maintain reliable healthcare systems is fundamental to achieving health equity. Despite numerous initiatives to combat health disparities, the current healthcare landscape in LMICs demonstrates that these countries remain far behind. With personalized medicine emerging as a new frontier in healthcare, its promising benefits may have serious implications, potentially making health inequity in developing countries irreconcilable. Multiple studies have already shown how resource-intensive personalized medicine is, and to date, its methods are primarily adopted in high-income countries. As a result, success stories of its implementation have not yet been fully explored in LMICs. Personalized medicine relies on advanced technology for genomics, diagnostics, and specialized treatments, which require significant investments in new technology and workforce training (Owolabi et al., 2023). While genomics has advanced greatly in wealthy countries with immediate infrastructure establishment, its application in low-resource countries remains premature (Mitropoulos et al., 2017).

According to Walters et al. (2023), most LMICs face major obstacles in integrating genetics into primary healthcare systems. These obstacles often include the absence or limited availability of genetic services, a shortage of skilled human resources, and inadequate facilities, tools, and technology for genetic testing. Furthermore, political and financial barriers, such as insufficient health policies and resources, place restrictions on accessible genetic services, passing financial burdens onto patients, who face high costs for these services. Finally, cultural and religious beliefs also limit access to and acceptance of genetic integration, often linked to healthcare providers' lack of understanding and knowledge of genetics.

Given these pervasive challenges, achieving health equity in the era of personalized medicine requires reimagined global health strategies. Foundational issues impacting health equity must be addressed to implement personalized medicine effectively in LMICs. This demands a comprehensive review of educational approaches, socio-political factors, legislative frameworks, and ethical considerations to overcome primary challenges before transitioning to this new healthcare paradigm. Key initiatives to help bridge this gap include promoting scientific education, particularly in computational methods, genomics, and genetics fundamentals, updating healthcare curricula, and launching health literacy campaigns. Additionally, revisiting and amending health policies to encourage collaborations and resource sharing is essential. In Thailand, for example, Thamlikitkul (2023) demonstrates how their government has established accessible public and private genetic testing services for cancer across all regions. Their progress in genomic medicine has been facilitated by substantial funding for cancer precision research. Similarly, global health initiatives are shifting toward frameworks for policy reforms in healthcare workforce development, equitable resource allocation, and ethical data sharing (World Health Organization, 2022; Ambrosino et al., 2024), all foundational steps for advancing genomics and implementing personalized medicine in LMICs.

While transitioning to personalized medicine is time-consuming and resource-intensive, addressing primary issues, from basic healthcare infrastructure to advanced technologies, can help LMICs establish a sustainable framework for this paradigm. In essence, though personalized medicine poses significant risks of exacerbating health disparities, it also creates an opportunity for transdisciplinary collaboration, especially when stakeholders from high-, middle-, and low-income countries join forces to tackle shared challenges. Therefore, it is imperative to consider ethical implications and strengthen efforts to make personalized medicine accessible to all, promoting equity in healthcare for the future.

**4. Innovative Approaches to Aligning Personalized Medicine with Health Equity**

Emerging technologies such as AI, specifically machine learning (ML) and federated learning, along with high-throughput screening, have streamlined the previously time-consuming process of precision medicine. AI and ML have been instrumental in identifying and analyzing patterns from patient data, significantly improving disease diagnosis, clinical testing, treatment personalization, and overall healthcare quality (Alowais et al., 2023; Parekh et al., 2023). As a result, AI integration in precision medicine has reduced costs by enhancing diagnostic accuracy and shortening the trial-and-error phase, all while delivering optimal patient care (Johnson et al., 2021). Federated learning, in particular, is pivotal for healthcare as it enables multiple hospitals to participate in AI model training while preserving data privacy (Rahman et al., 2022). This approach centers communication on the model itself without requiring data exchange, allowing devices to train or update models while keeping data locally stored, thereby reducing the risk of data disclosure. Incorporating both AI and federated learning into precision medicine can enhance healthcare access and delivery while promoting collaborative efforts.

High-throughput sequencing, widely used in the pharmaceutical industry, evaluates the biological activity of chemicals against specific targets for drug discovery (Kolukisaoglu and Thurow, 2010). Recent advancements in sequencing technologies have made high-throughput sequencing even more valuable in developing personalized medicine (Saxena et al., 2022). For example, high-throughput screening has been applied to analyze tumor samples from 125 pediatric patients, identifying their activated genomic alterations and responses to drug treatment (Mayoh et al., 2023). Integrating high-throughput screening into precision medicine has improved biomarker-driven strategies for treating cancer, while automated testing has reduced costs compared to manual, labor-intensive methods, streamlining the drug development process.

Decentralized healthcare models enhance the quality and effectiveness of localized patient care by adopting patient-centered approaches within the community. Mobile health, or mHealth, is one such model that delivers healthcare through telecommunications on smartphones and wearable devices (Pahlevanynejad et al., 2023). These portable devices enable healthcare professionals to provide services despite time and geographical barriers. Remote monitoring of patients is facilitated through mHealth in personalized medicine, allowing physicians to access clinical decision support based on device data (Hilty et al., 2019). Continuous monitoring via mHealth supports rapid disease diagnosis, prevention, and management, reducing medical errors and overall costs.

Telemedicine has substantially grown during the COVID-19 pandemic, providing essential healthcare services through call and video conferencing platforms. In the Philippines, telemedicine became a vital healthcare access point amid pandemic restrictions, with the Department of Health promoting it as part of healthcare in the "new normal" (Macariola et al., 2021). Telemedicine addresses immediate healthcare needs while supporting long-term goals for decentralized healthcare. Through customized diagnostic and therapeutic approaches, telemedicine supports personalized medicine and helps physicians gain a holistic understanding of each patient's life circumstances (Record et al., 2021). Studies report high satisfaction rates with telemedicine, largely due to its focus on patient-centered communication, enhanced accessibility, and the involvement of family members or support individuals during teleconsultations (Orlando et al., 2019; Record et al., 2021; Smith et al., 2022). This shift has been a significant step toward personalized medicine, as telemedicine reduces hospital visits, encourages patients to manage their conditions, lowers transport costs, and decreases the risk of hospital-acquired infections (Pilosof et al., 2021; Mudiyanselage et al., 2023).

Despite these benefits, there are barriers to implementing telemedicine. Issues like low connectivity, technology illiteracy, and lack of access to mobile devices directly impact telemedicine's effectiveness and can contribute to healthcare disparities (Centers for Disease Control and Prevention, 2020). Addressing these limitations by improving technological literacy and access to mobile devices could enhance telemedicine's reliability and accessibility, especially in low-resource communities. Moving forward, integrating telemedicine into personalized medicine will significantly improve healthcare accessibility.

Finally, the World Health Organization's (WHO) Global Health Initiative (GHI) has been instrumental in addressing emerging diseases, malnutrition, maternal and child health, and immunization programs in low-income countries. Through GHI programs, there has been a substantial decrease in high-priority diseases such as malaria, HIV, and vaccine-preventable diseases (Nishio, 2023). GHI, in partnership with Gavi, provides affordable healthcare to low-income countries. Gavi's vaccine alliance, through public-private partnerships and multilateral funding, negotiates lower vaccine prices, making them accessible to vulnerable populations and low-income countries (KFF, 2024). By 2020, Gavi had introduced 200 new vaccine programs to low-income countries, improving vaccination rates among children and expectant mothers amid rising demand (Zerhouni, 2019). The impact of Gavi on public health through its vaccination initiatives and lives saved is profound.

## 5. Ethical Considerations in the Push for Personalization

The pursuit of personalized medicine raises several ethical dilemmas, particularly regarding unequal access to cutting-edge treatments. Personalized therapies often come with high costs, limited availability, and a need for sophisticated infrastructure, which risks concentrating the benefits of these innovations in high-resource settings. This leaves low-resource countries with inadequate healthcare options, exacerbating health disparities between wealthy economies and LMICs. This divide raises questions about the fairness of personalized treatment applications in the global context and highlights the need to explore the theoretical foundations of these ethical dilemmas to assess the tension between personalized medicine and health equity.

One of the primary ethical dilemmas is the unequal access to personalized treatments between populations with different economic standings. Treatments like gene therapies, tailored pharmaceuticals, and personalized diagnostics often require extensive financial resources and advanced healthcare infrastructure, which are typically lacking in LMICs. As a result, affluent populations have priority access to life-saving treatments, while low-resource populations are left with inadequate healthcare options.

Various classical and ethical theories offer unique perspectives on the distribution of healthcare resources. From a utilitarian perspective, healthcare policies should aim to maximize overall well-being. Advocates of this view might argue that personalized medicine, though expensive, could still be justified if it results in significant health gains for the greatest number of people (Gillon et al., 1985; Mandal et al., 2016). However, a utilitarian approach may also prioritize treatments that benefit a larger number of individuals in low-resource settings with common, preventable diseases rather than focusing on costly, niche treatments for individuals in high-resource settings. Marseille and Khan (2019) note that, while utilitarianism has its limitations, it values cost-effectiveness and often supports achieving the most health benefits without exceeding available financial resources. In this context, utilitarianism would prioritize solutions that produce the greatest health impact, such as focusing on public health infrastructure before personalized medicine.

On the other hand, Rawls' theory of justice, specifically the "difference principle" (Rawls, 1971), challenges the ethical justification of any healthcare innovation that worsens existing inequalities. This theory posits that inequalities are acceptable only if they benefit the least advantaged members of society (Drane, 1990). In this case, developing and implementing personalized medicine in a way that risks exacerbating health inequalities would be unethical. From a Rawlsian standpoint, personalized medicine must be equally accessible to all, with priority given to disadvantaged communities before benefiting those in high-resource settings. Kantian ethics on universality and respect for persons (Kant, 1785) similarly critiques healthcare systems that offer specialized treatments only to those who can afford them. This approach implies a moral obligation for healthcare systems to ensure equitable access, respecting the dignity and rights of every individual (Merritt et al., 2007).

Immanuel Kant's ethical theory aligns closely with the capabilities approach proposed by philosophers Martha Nussbaum and Amartya Sen. This theory emphasizes the importance of providing individuals with the opportunity to fulfill their potential (Crocker, 1995). In the context of personalized medicine, this theory would argue that access to healthcare technologies should not depend on economic factors or resource availability. Instead, governments and institutions should foster environments that enable everyone to achieve their full potential on a global scale. This approach advocates for equal opportunities for all individuals to benefit from personalized medicine, encouraging global health policies to be reformed to promote health equity and well-being.

Modern theories of distributive justice, as developed by Norman Daniels, emphasize the ethical imperative to ensure "fair equality of opportunity" in health (Daniels, 2008). This theory asserts that health disparities that prevent individuals from fully participating in society are unjust. In the context of personalized medicine, this ethical framework challenges any approach that allows technological advances to widen the gap between vulnerable low-resource populations and high-income groups. It argues that access to personalized therapies and treatments should be distributed equally to promote equitable health opportunities. This perspective supports the idea that technology should be optimized to improve health outcomes for all (Saeed & Masters, 2021).

These ethical considerations are particularly relevant to address whether personalized healthcare should be prioritized in high-resource economies first. Proponents of this approach argue that these regions can serve as testing grounds for optimizing health treatments due to their resource availability. However, this approach is ethically problematic, as it risks fostering "technological elitism" and widening global health gaps, creating a "biomedical apartheid" that gives high-income populations greater advantages in accessing these technologies over poorer countries. Moreover, Goodman and Brett (2018) emphasize that genetic diversity within certain geographic or ethnic groups is more pronounced than between racial categories. Hussein et al. (2022) argue that applying personalized medicine in regions like Africa requires a more nuanced approach due to the continent's vast genetic diversity and environmental variation. This view is corroborated by Drake et al. (2018), who highlights that genomic advances in high-resource countries cannot represent the entire global population.

From a utilitarian perspective, prioritizing high-resource settings might still be justified if the long-term goal is to create scalable, cost-effective treatments that can eventually be applied globally. However, a Rawlsian or capabilities-based approach would demand that healthcare systems ensure equitable access from the outset, enabling the least advantaged populations to benefit from these technologies simultaneously with wealthier populations. This highlights the need for existing governing bodies, global health leaders, and institutions to provide an ethical framework that emphasizes fair distribution and commits to mobilizing strategies that promote health equity in personalized medicine.

A critical reassessment of how to manage current ethical dilemmas and healthcare advancements in line with justice, equity, and universal human dignity is essential for all stakeholders involved. This calls for strategic efforts to mitigate existing health disparities and reinforce health policies that balance the distribution of personalized medical interventions. Strengthening collaborations can support the development of personalized medicine that serves individuals across socio-economic backgrounds, paving the way for a more inclusive and just healthcare system in the future.

## 6. Case Studies: Successes and Shortcomings

High-income countries are classified by the World Bank based on their gross national income (GNI) per capita, which is typically around $14,005 USD, whereas lower-middle-income countries have a GNI per capita ranging from $1,146 USD to $4,515 USD (World Bank Country and Lending Groups, n.d). Australia, the USA, Japan, France, and Germany are examples of these high-income countries. Through ongoing genomic testing efforts, personalized medicine is widely available to citizens in these countries. Implementation efforts are further supported by government funding, which provides access to accurate genetic diagnoses, additional genomic screening, and clinical management support for families (O'Shea et al., 2022). The United Kingdom is a pioneer in implementing personalized cancer medicine, tailoring treatments to patients through extensive genomic profiling, molecular pathology, and AI to analyze large datasets from cancer patients accurately and efficiently (Masucci et al., 2024). Similarly, in the US, personalized medicine has been applied in cardiac care, using technologies like metabolomics, proteomics, and microbiomics to deliver accurate diagnoses and customized treatments for cardiovascular disease patients (Sethi et al., 2023). These examples demonstrate that personalized medicine effectively ensures high-quality healthcare through novel technologies and tailored treatments.

In low- and middle-income countries (LMICs), implementing personalized medicine is challenging due to a lack of financial support. Additional barriers include limited genomic data from the population, restricted access to diagnostic tools, limited availability and affordability of specific drugs, and insufficient infrastructure to support personalized medicine (Gameiro et al., 2018; Thomas et al., 2023). Despite these obstacles, several initiatives have successfully integrated personalized medicine in LMICs in a cost-effective manner. For instance, the PerMediNA initiative, launched in North African countries including Tunisia, Algeria, and Morocco, is implementing a pilot oncology precision project supported by healthcare professionals on the molecular tumor board (Hamdi et al., 2024). This project aims to streamline the discovery of novel treatments for genetic targets and develop customized treatments for each cancer patient while minimizing adverse effects. Additionally, it seeks to translate research findings into cost-effective and sustainable health services.

Another successful initiative is the partnership between the Rwandan government and BGI Genomics to launch over 20,000 cervical cancer testing kits for Human Papilloma Virus (HPV). Screening efforts in Rwanda have been strengthened to address the rising risk of cervical cancer, with approximately 3.7 million middle-aged women at risk of developing the disease as of 2020 (Gafaranga et al., 2022). This partnership between the government and the private sector aims to enhance cervical cancer screening and prevention for women in Rwanda while advancing local precision medical testing capabilities (Ecancer, 2020). This initiative demonstrates the potential of public-private partnerships to improve health outcomes by making personalized medicine accessible and sustainable, especially in LMICs.

Highlighting the disparity in personalized medicine between high-income and low- and middle-income countries is essential to achieving health equity. A comprehensive approach addressing the social, economic, and political factors that impact the healthcare system is needed. Additionally, opportunities such as external research funding and government collaboration with non-profits or private corporations could jumpstart the implementation of personalized medicine in LMICs, providing them with funds, healthcare worker training, and access to advanced diagnostic technologies. Through collaborative efforts and innovative partnerships, it is possible to implement personalized medicine even in low-income countries, ultimately improving health outcomes and access to new therapeutics.

**7. Proposals for Policy and Structural Changes**

To address the disparity between personalized medicine and health equity, policy changes are necessary to ensure that healthcare services offered by personalized medicine are accessible and affordable, especially for LMICs. Bridging the gaps between current health inequities and the potential of personalized medicine requires a collective effort from socio-political stakeholders and healthcare professionals. Existing global health initiatives can be expanded and improved to better address and adapt to local needs and resources, supporting equitable access to personalized care.

For example, the International Consortium for Personalized Medicine (ICPerMed) is composed of experts who aim to foster initiatives focused on healthcare system improvement, medication market access, and patient empowerment (Venne et al., 2020). Since its launch in 2016, the consortium has created and developed numerous action plans, frameworks, and funding roadmaps designed to support both local and international research, education, and innovative solutions. This collaboration across different sectors promotes and helps accelerate the development and implementation of personalized medicine. In 2020, ICPerMed outlined five key perspectives as a framework to guide personalized medicine initiatives over the next 10 years with the goal of achieving both personalized medicine and health equity. These perspectives include promoting individual and community mobility, engaging the healthcare workforce actively, integrating personalized medicine practices, expanding health-related data, and ensuring economic sustainability to support improved personalized medicine approaches (Vicente et al., 2020).

ICPerMed also suggests specific strategies, such as developing IT solutions for big data collection, management, and sharing, as well as providing comprehensive education and skills training for everyone, including healthcare providers. According to Pastorino et al. (2019), the collection of patient data in healthcare is the first step toward improving disease prevention and patient care quality. Big data, encompassing patient information from electronic health records, genomic testing results, prescriptions, and imaging results, is now an integral part of personalized medicine, helping improve efficiency, reduce diagnostic errors, and lower treatment costs (Badr et al., 2024). To execute this vision and overcome the barriers between personalized medicine and health equity, the following recommendations should be considered:

### *7.1. Educational Reform*

Personalized medicine, especially in LMICs, is in its infancy and remains largely underdeveloped in some regions. One of the most significant barriers to its adoption, aside from cost, is the considerable lack of awareness, training, and education among healthcare providers and the general public. Targeted curriculum development is essential for building a community prepared to address shifts in the healthcare landscape and improve existing conditions.

Currently, strong hesitancy among medical practitioners regarding the adoption of personalized medicine is evident. Often overlooked in traditional medical education, recent literature notes that only 21% of discussions on the principles of personalized medicine are included in medical school curricula (Lamichhane & Agrawal, 2023). This likely contributes to the lack of awareness and doubts surrounding the application of personalized medicine in clinical practice. Moving forward, reforming medical curricula to introduce the principles and concepts of personalized medicine is essential. Additionally, other allied health programs play a crucial role and must also incorporate education on the effective application of personalized medicine into their focus and practices. Within this evolving landscape, healthcare workers must be able to meet the demands of the modern healthcare system. Spanakis et al. (2020) suggest that nursing education should adapt to the changing healthcare environment by advancing expertise in genomics, mathematics, statistics, ethics, and information and communication technologies. This approach will ensure that safe, well-informed, and effective personalized nursing care can be delivered. Similarly, introducing these transdisciplinary foundational courses at an early stage in nursing education will support the expansion of nursing roles in personalized healthcare.

Academic programs can be designed and enhanced to incorporate the principles of personalized medicine, providing a strong foundation in genetics and genomics, data analysis, and practical application to develop a well-rounded and skilled workforce capable of utilizing personalized medicine and optimizing healthcare outcomes. It is important to acknowledge that adopting personalized medicine requires a unified approach that merges various fields such as genetics, bioengineering, and information technology. Unlike traditional practices, personalized medicine demands collaboration across diverse disciplines (Sun et al., 2018). This need for synergy is exemplified by Yale's new graduate program, launched in 2022, offering a Master of Science in Personalized Medicine and Applied Engineering[1]. This program reflects the necessity of building capacity among the next generation of healthcare and engineering professionals focused on improving personalized patient care. Such collaboration enhances precise treatments, provides a comprehensive understanding of patient needs, and fosters a healthcare infrastructure built on teamwork, with a vision to strengthen the modern global healthcare system.

Finally, implementing health literacy campaigns to promote public awareness can also improve perceptions of personalized medicine. These campaigns encourage community engagement and openness to professional healthcare, helping remote communities understand and reduce misconceptions about medical interventions, ultimately promoting better health (Trein & Wagner, 2021).

### 7.2. International Collaboration

In promoting global health equity, international collaborations are essential. Establishing strong relationships between governments, experts, and international organizations is recommended to accelerate equitable access to personalized medicine. Multilateral agreements between developed and developing countries can foster inclusive progress through resource sharing in technology, knowledge, and innovations. Global initiatives like ICPerMed illustrate this potential by creating a platform for dialogue to drive advancements in personalized healthcare. Additionally, partnerships between countries that have already integrated personalized medicine into their healthcare systems can support those still facing adoption challenges through exchange programs and training, where both sides benefit from gaining insights into potential challenges in personalized medicine and how to address them efficiently.

### 7.3. Resource allocation and Research and Development

For many countries, the cost of personalized medicine is the most challenging barrier to overcome. Addressing this challenge requires a comprehensive approach. While the long-term solution involves economic evaluation and a reassessment of priorities in government resource allocation, immediate actions are also essential.

Governments are encouraged to prioritize funding for the healthcare sector, along with research and development, to improve strained healthcare systems, particularly in LMICs. Partnerships with private sectors are equally important to accelerate innovative solutions. Industry stakeholders, policymakers, and other governing bodies serve as primary facilitators who can mobilize and ease the implementation of personalized medicine. The health and well-being of the community must, and should always, be the top priority. Existing global health initiatives from international organizations, such as ICPerMed and the World Health Organization, offer frameworks that can serve as practical guidelines for governments and other key stakeholders to take action. These initiatives can help build a more resilient healthcare system and achieve health equity.

---

[1] Yale School of Medicine. (2022). MS in Personalized Medicine & Applied Engineering. Medicine.yale.edu. Retrieved from https://medicine.yale.edu/ortho/education/masters-personalized-medicine-applied-engineering/

*7.4. Maximizing New Technology*

Technology has become an indispensable tool for positively transforming healthcare and addressing significant challenges and barriers (Fermin and Tan, 2020). The hallmarks of personalized medicine rely on technological innovations and advancements. Tools like artificial intelligence and machine learning help streamline the work of healthcare providers, easing the considerable burden this new approach places on an already strained healthcare workforce. Effective data utilization is crucial in personalized medicine, but concerns over data security persist among various stakeholders (Cascini et al., 2024; Zhang et al., 2024). To address these concerns while maximizing technology's potential, federated learning offers a privacy-preserving approach. Rather than sharing raw data, federated learning allows healthcare systems to build and refine predictive models collaboratively by processing data locally on devices, ensuring that sensitive patient information remains secure and never leaves its original location (Dhade & Shirke, 2024).

More recently, technology has effectively expanded the reach of healthcare services. Telemedicine and mobile health (mHealth) exemplify how technology can bridge personalized medicine and health equity by facilitating accessible interactions between providers and patients. This approach enables people from underserved locations and the broader community to access specialized care more easily. As technology continues to evolve, harnessing its full potential can significantly enhance the inclusivity of personalized medicine, promoting better health outcomes across diverse populations.

Bridging the gap between health equity and personalized medicine requires a multi-faceted approach, involving policy reform, educational development, international collaboration, and strategic technology use. Socio-political stakeholders must prioritize equitable access to personalized medicine, especially in low-resource populations where health disparities are often more pronounced. Building on global health initiatives like ICPerMed fosters collaborative efforts that support patient empowerment and sustainable healthcare practices. Reforms in educational curricula and government legislation promote learning, innovation, and more accessible healthcare. Partnerships across the globe are also essential; building meaningful relationships can enhance and facilitate the implementation of personalized medicine through resource sharing. Finally, the strategic use of technology like federated learning can streamline complex processes, reduce healthcare system strain, protect patient data privacy, and broaden healthcare access by removing geographical barriers to equitable care.

These concerted efforts can substantially ease the gradual removal of existing barriers to health, building capacity for a more resilient and accessible healthcare system. With full commitment and consistency in implementing these strategies, the healthcare landscape can transform, making personalized medicine and health equity a reality for all, regardless of socio-economic status or geographical location.

**8. Conclusion: Moving Toward a Balanced Future**

Promising new paradigms in healthcare lose their value when they cannot serve all individuals equitably. The future of personalized medicine lies in its potential to improve quality of life through a transformed healthcare system, but first, it must address a critical question: can these advancements coexist with a commitment to health equity? This paper has highlighted several challenges hindering the full realization of personalized medicine and the risks it poses in potentially increasing health disparities. LMICs, in particular, face numerous healthcare barriers, and the added demands of personalized medicine—such as high costs, robust infrastructure needs, and capacity requirements—further complicate accessibility. Without intervention, personalized medicine may remain an advantage for countries with strong economies, leaving LMICs to grapple with an even greater healthcare equity gap.

To address this issue, transdisciplinary stakeholders must make decisions to ensure that personalized medicine is implemented with health equity at its core. Figure 1 illustrates transdisciplinary approaches to establishing personalized medicine in healthcare with a central focus on health equity. Achieving this goal requires concerted efforts from policymakers, industry, world leaders, and healthcare providers to mobilize and strategically develop sustainable solutions that create a more inclusive landscape for personalized medicine adoption. These efforts include significant financial investments, strong partnerships, and ethical frameworks to enhance the capacity of LMICs. Expanding education and training, leveraging innovative technologies, and enacting legislation to maintain accessible and affordable personalized medicine services will also help mitigate disparities while maximizing the benefits of this healthcare advancement. Achieving health equity within personalized medicine requires meaningful collaboration and sustained efforts from all stakeholders.

While this paper centers on health equity within personalized medicine, it is important to acknowledge existing barriers that continue to obstruct low-resource populations from accessing healthcare. Often overlooked in favor of new innovations, these challenges must be recognized and addressed. If current issues are not resolved, an already strained healthcare system will be unable to integrate personalized medicine effectively, calling for immediate action from global stakeholders. Future studies must focus on practical and innovative solutions that address the financial and infrastructural limitations faced by LMICs. Developing frameworks to make personalized medicine sustainable and more affordable is essential. Health equity can then be achieved, putting the vision of improved health for everyone within reach.

One such innovation is federated learning, which enables multiple healthcare providers to develop AI models without sharing sensitive patient data. Federated learning provides a scalable way for LMICs to collaborate with high-resource countries on critical research and AI model development while safeguarding data privacy and reducing infrastructure demands. By ensuring all individuals can benefit from advancements in personalized medicine, federated learning exemplifies how innovative technology can foster an inclusive healthcare system. With unified efforts toward this common goal, the benefits of personalized medicine can be fully realized across borders, regardless of socio-economic status, paving the way for a healthcare landscape where innovation and equity coexist.

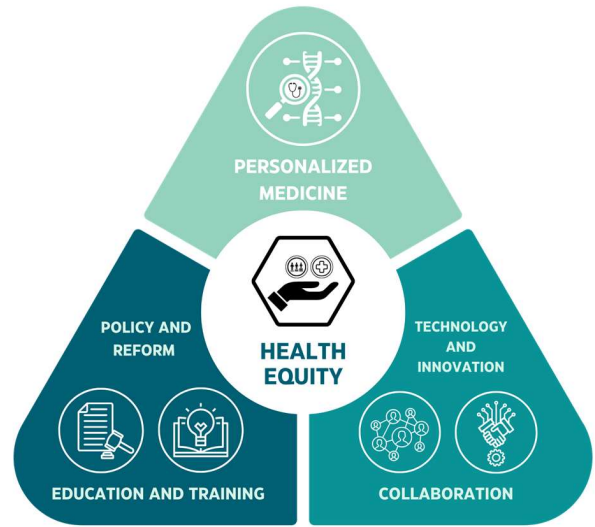

**Figure 1:** *Transdisciplinary framework for equitable personalized medicine.* This transdisciplinary framework for Health Equity in Personalized Medicine positions Health Equity as the core guiding principle for the realization of equitable personalized healthcare. The model divides its five interconnected pillars into foundational and practical elements essential for building an inclusive and resilient healthcare system. On the left side, Policy and Reform and Education and Training represent foundational elements that equip healthcare providers and other stakeholders with the necessary knowledge, awareness, and skills to implement accessible personalized medicine. Policy reforms help remove financial and regulatory barriers, while targeted education develops a skilled workforce and enhances public health literacy. On the right side, Technology and Innovation and Collaboration offer practical solutions to directly support the delivery and accessibility of personalized medicine, addressing ongoing challenges in health equity through advanced tools like AI and big data, alongside resource-sharing partnerships between governments, industry, and organizations. At the top, Personalized Medicine integrates these foundational and practical elements to ensure precision diagnostics and treatments reach diverse populations. Together, these five pillars create a comprehensive, transdisciplinary approach to align personalized medicine with health equity, fostering an inclusive and sustainable global healthcare system.

**References**

1. Adeniji AA, Dulal S, Martin MG. Personalized Medicine in Oncology in the Developing World: Barriers and Concepts to Improve Status Quo. World J Oncol. 2021 Jun;12(2-3):50-60. doi: 10.14740/wjon1345. Epub 2021 May 14. PMID: 34046099; PMCID: PMC8139741.

2. Agyeman, A. A., & Ofori-Asenso, R. (2015). Perspective: Does personalized medicine hold the future for medicine?. Journal of pharmacy & bioallied sciences, 7(3), 239–244. https://doi.org/10.4103/0975-7406.160040

3. Alibudbud, R. (2023). Addressing the burnout and shortage of nurses in the Philippines. SAGE Open Nursing, 9, 23779608231195737.

4. Alowais, S. A., Alghamdi, S. S., Alsuhebany, N., Alqahtani, T., Alshaya, A. I., Almohareb, S. N., Aldairem, A., Alrashed, M., Bin Saleh, K., Badreldin, H. A., Al Yami, M. S., Al Harbi, S., & Albekairy, A. M. (2023). Revolutionizing healthcare: the role of artificial intelligence in clinical practice. *BMC medical education*, *23*(1), 689. https://doi.org/10.1186/s12909-023-04698-z

5. Alyass, A., Turcotte, M., & Meyre, D. (2015). From big data analysis to personalized medicine for all: challenges and opportunities. *BMC medical genomics*, *8*, 1-12.

6. Ambrosino, E., Abou Tayoun, A.N., Abramowicz, M. et al. The WHO genomics program of work for equitable implementation of human genomics for global health. Nat Med 30, 2711–2713 (2024). https://doi.org/10.1038/s41591-024-03225-x

7. Bayati, M., Noroozi, R., Ghanbari-Jahromi, M., & Jalali, F. S. (2022). Inequality in the distribution of Covid-19 vaccine: a systematic review. *International journal for equity in health*, *21*(1), 122.

8. Biglan, A., Prinz, R. J., & Fishbein, D. (2023). Prevention science and health equity: A comprehensive framework for preventing health inequities and disparities associated with race, ethnicity, and social class. *Prevention Science*, *24*(4), 602-612.

9. Braveman, P. A., Kumanyika, S., Fielding, J., LaVeist, T., Borrell, L. N., Manderscheid, R., & Troutman, A. (2011). Health disparities and health equity: the issue is justice. *American journal of public health*, *101*(S1), S149-S155.

10. Braveman, P., Arkin, E., Orleans, T., Proctor, D., Acker, J., & Plough, A. (2018). What is Health Equity? *Behavioral Science & Policy*, *4*(1), 1–14. https://doi.org/10.1177/237946151800400102

11. Brittain, H. K., Scott, R., & Thomas, E. (2017). The rise of the genome and personalised medicine. *Clinical Medicine*, *17*(6), 545–551. https://doi.org/10.7861/clinmedicine.17-6-545

12. Burr, W. H., Everitt, J. G., & Johnson, J. M. (2023). "The debt is suffocating to be honest": Student loan debt, prospective sensemaking, and the social psychology of precarity in an allopathic medical school. SSM-Qualitative Research in Health, 4, 100304.

13. Canadian Agency for Drugs and Technologies in Health. (2023, July 1). *Trastuzumab Deruxtecan (Enhertu)*. NCBI Bookshelf. https://www.ncbi.nlm.nih.gov/books/NBK595132/

14. Cascini, F., Pantovic, A., Al-Ajlouni, Y. A., Puleo, V., De Maio, L., & Ricciardi, W. (2024). Health data sharing attitudes towards primary and secondary use of data: a systematic review. *EClinicalMedicine*, *71*, 102551. https://doi.org/10.1016/j.eclinm.2024.102551

15. Center for Disease Control and Prevention (2020). Using Telehealth to Expand Access to Essential Health Services During the COVID-19 Pandemic. https://www.cdc.gov/coronavirus/2019-ncov/hcp/telehealth.html

16. Choi, B. C., & Pak, A. W. (2006). Multidisciplinarity, interdisciplinarity and transdisciplinarity in health research, services, education and policy: 1. Definitions, objectives, and evidence of effectiveness. *Clinical and investigative medicine. Medecine clinique et experimentale*, *29*(6), 351–364.

17. Chong, H. Y., Allotey, P. A., & Chaiyakunapruk, N. (2018). Current landscape of personalized medicine adoption and implementation in Southeast Asia. *BMC Medical Genomics*, *11*(1). https://doi.org/10.1186/s12920-018-0420-4

18. Crocker, D. A. (1992). Functioning and capability: the foundations of Sen's and Nussbaum's development ethic. *Political theory*, *20*(4), 584-612.

19. Daniels, N. (2008). Justice and access to health care.

20. Deminco, F., Vaz, S., Santana, D., Pedroso, C., Tadeu, J., Stoecker, A., . . . Brites, C. (2022, October 27). A simplified Sanger sequencing method for detection of relevant SARS-COV-2 variants. https://www.mdpi.com/2075-4418/12/11/2609

21. De Vries, N., Boone, A., Godderis, L., Bouman, J., Szemik, S., Matranga, D., & De Winter, P. (2023). The race to retain healthcare workers: a systematic review on factors that impact retention of nurses and physicians in hospitals. *INQUIRY: The Journal of Health Care Organization, Provision, and Financing*, *60*, 00469580231159318.

22. Dhade, P., & Shirke, P. (2024). Federated Learning for Healthcare: A Comprehensive Review. *Engineering Proceedings*, *59*(1), 230.

23. Drake, T. M., Knight, S. R., Harrison, E. M., & Søreide, K. (2018). Global inequities in precision medicine and molecular cancer research. *Frontiers in Oncology*, *8*, 346.

24. Drane, J. F. (1990). Justice issues in health care delivery. *Bulletin of the Pan American Health Organization (PAHO); 24 (4), 1990*.

25. Diaby, V., Alqhtani, H., Van Boemmel-Wegmann, S., Wang, C., Ali, A. A., Balkrishnan, R., Ko, Y., Palacio, S., & De Lima Lopes, G. (2019). A cost-effectiveness analysis of trastuzumab-containing treatment sequences for HER-2 positive metastatic breast cancer patients in Taiwan. *The Breast*, *49*, 141–148. https://doi.org/10.1016/j.breast.2019.11.012

26. Duroseau, B., Kipshidze, N., & Limaye, R. J. (2023). The impact of delayed access to COVID-19 vaccines in low-and lower-middle-income countries. *Frontiers in public health*, *10*, 1087138.

27. Eaton, J., Baingana, F., Abdulaziz, M., Obindo, T., Skuse, D., & Jenkins, R. (2023). The negative impact of global health worker migration, and how it can be addressed. *Public health*, *225*, 254-257.

28. Ecancer. (2024, May 13). Rwanda initiative: public health boost with cervical cancer screening for 20,000. *Ecancer*. https://ecancer.org/en/news/24726-rwanda-initiative-public-health-boost-with-cervical-cancer-screening-for-20-000

29. Fermin, J. L., & Tan, M. J. T. (2020). The need for the establishment of biomedical engineering as an academic and professional discipline in the Philippines—A quantitative argument. *IEEE Access*, *9*, 3097-3111.

30. Fröhlich, H., Balling, R., Beerenwinkel, N., Kohlbacher, O., Kumar, S., Lengauer, T., Maathuis, M. H., Moreau, Y., Murphy, S. A., Przytycka, T. M., Rebhan, M., Röst, H., Schuppert, A., Schwab, M., Spang, R., Stekhoven, D., Sun, J., Weber, A., Ziemek, D., & Zupan, B. (2018). From hype to reality: data science enabling personalized medicine. *BMC Medicine*, *16*(1). https://doi.org/10.1186/s12916-018-1122-7

31. Gafaranga, J. P., Manirakiza, F., Ndagijimana, E., Urimubabo, J. C., Karenzi, I. D., Muhawenayo, E., Gashugi, P. M., Nyirasebura, D., & Rugwizangoga, B. (2022). Knowledge, Barriers and Motivators to Cervical cancer screening in Rwanda: a Qualitative study. *International Journal of Women S Health*, *Volume 14*, 1191–1200. https://doi.org/10.2147/ijwh.s374487

32. Gameiro, G. R., Sinkunas, V., Liguori, G. R., & Auler-Júnior, J. O. C. (2018). Precision Medicine: Changing the way we think about healthcare. *Clinics*, *73*, e723. https://doi.org/10.6061/clinics/2017/e723

33. Genuino, A. J., Chaikledkaew, U., Guerrero, A. M., Reungwetwattana, T., & Thakkinstian, A. (2019). Cost-utility analysis of adjuvant trastuzumab therapy for HER2-positive early-stage breast cancer in the Philippines. *BMC Health Services Research*, *19*(1). https://doi.org/10.1186/s12913-019-4715-8

34. Grant-Smith, D., & de Zwaan, L. (2019). Don't spend, eat less, save more: Responses to the financial stress experienced by nursing students during unpaid clinical placements. Nurse education in practice, 35, 1-6.

35. Goodman, C. W., & Brett, A. S. (2021). Race and pharmacogenomics—personalized medicine or misguided practice?. *Jama*, *325*(7), 625-626.

36. Hamdi, Y., Boujemaa, M., Aissa-Haj, J. B., Radouani, F., Khyatti, M., Mighri, N., ... & Abdelhak, S. (2024). A regionally based precision medicine implementation initiative in North Africa: The PerMediNA consortium. *Translational Oncology*, *44*, 101940.

37. Hirsch, B. R., & Schulman, K. A. (2013). The economics of new drugs: Can we afford to make progress in a common disease? *American Society of Clinical Oncology Educational Book*, *33*, e126–e130. https://doi.org/10.14694/edbook_am.2013.33.e126

38. Hussein, A. A., Hamad, R., Newport, M. J., & Ibrahim, M. E. (2022). Individualized medicine in Africa: bringing the practice into the realms of population heterogeneity. Frontiers in Genetics, 13, 853969.

39. Ikhurionan, P., Kwarshak, Y. K., Agho, E. T., Akhirevbulu, I. C., Atat, J., Erhiawarie, F., ... & Wariri, O. (2022). Understanding the trends, and drivers of emigration, migration intention and non-migration of health workers from low-income and middle-income countries: protocol for a systematic review. BMJ open, 12(12), e068522

40. Jakka, S., & Rossbach, M. (2013). An economic perspective on personalized medicine. *The HUGO Journal*, *7*(1). https://doi.org/10.1186/1877-6566-7-1

41. Jensen, N., Kelly, A. H., & Avendano, M. (2021). The COVID-19 pandemic underscores the need for an equity-focused global health agenda. *Humanities and Social Sciences Communications*, *8*(1).

42. Johnson, K. B., Wei, W. Q., Weeraratne, D., Frisse, M. E., Misulis, K., Rhee, K., Zhao, J., & Snowdon, J. L. (2021). Precision Medicine, AI, and the Future of Personalized Health Care. *Clinical and translational science*, *14*(1), 86–93. https://doi.org/10.1111/cts.12884

43. Kant, Immanuel (1785). Groundwork for the metaphysics of morals. New York: Oxford University Press. Edited by Thomas E. Hill & Arnulf Zweig.

44. KFF. (2024, July 17). *The U.S. government and GAVI, the Vaccine Alliance | KFF*. https://www.kff.org/global-health-policy/fact-sheet/the-u-s-government-gavi-the-vaccine-alliance/

45. Khorram-Manesh, A., Goniewicz, K., & Burkle, F. M. (2023). Social and Healthcare Impacts of the Russian-Led Hybrid War in Ukraine – A Conflict With Unique Global Consequences. Disaster Medicine and Public Health Preparedness, 17, e432. doi:10.1017/dmp.2023.91

46. Kristin, E., Endarti, D., Khoe, L., Pratiwi, W., Pinzon, R., Febrinasari, R., Yasmina, A., Nugrahaningsih, D. a. A., Taroeno-Hariadi, K., Karsono, R., Sudarsa, I. W., Prenggono, M., Herlinawaty, E., Komaryani, K., Hidayat, B., & Nadjib, M. (2022). Does Trastuzumab Offer Good Value for Money for Breast Cancer Patients with Metastasis in Indonesia? *Asian Pacific Journal of Cancer Prevention*, *23*(7), 2441–2447. https://doi.org/10.31557/apjcp.2022.23.7.2441

47. Lamichhane, P., & Agrawal, A. (2023). Precision medicine and implications in medical education. *Annals of medicine and surgery (2012)*, *85*(4), 1342–1345. https://doi.org/10.1097/MS9.0000000000000298

48. Leighl, N. B., Nirmalakumar, S., Ezeife, D. A., & Gyawali, B. (2021). An arm and a leg: the rising cost of cancer drugs and impact on access. *American Society of Clinical Oncology Educational Book*, *41*, e1–e12. https://doi.org/10.1200/edbk_100028

49. Macariola, A. D., Santarin, T. M. C., Villaflor, F. J. M., Villaluna, L. M. G., Yonzon, R. S. L., Fermin, J. L., Kee, S. L., AlDahoul, N., Karim, H. A., & Tan, M. J. T. (2021). Breaking Barriers Amid the Pandemic: The Status of Telehealth in Southeast Asia and its Potential as a Mode of Healthcare

Delivery in the Philippines. *Frontiers in Pharmacology*, *12*. https://doi.org/10.3389/fphar.2021.754011

50. Mandal, J., Ponnambath, D. K., & Parija, S. C. (2016). Utilitarian and deontological ethics in medicine. *Tropical parasitology*, *6*(1), 5–7. https://doi.org/10.4103/2229-5070.175024

51. Marseille, E., & Kahn, J. G. (2019). Utilitarianism and the ethical foundations of cost-effectiveness analysis in resource allocation for global health. *Philosophy, Ethics, and Humanities in Medicine*, *14*(1), 5.

52. Masucci, M., Karlsson, C., Blomqvist, L., & Ernberg, I. (2024). Bridging the Divide: A review on the implementation of personalized cancer medicine. *Journal of Personalized Medicine*, *14*(6), 561. https://doi.org/10.3390/jpm14060561

53. Mathur, S., & Sutton, J. (2017). Personalized medicine could transform healthcare. *Biomedical Reports*, *7*(1), 3–5. https://doi.org/10.3892/br.2017.922

54. Meckley, L. M., & Neumann, P. J. (2009). Personalized medicine: Factors influencing reimbursement. *Health Policy*, *94*(2), 91–100. https://doi.org/10.1016/j.healthpol.2009.09.006

55. Merritt, M. (2007). Bioethics, philosophy, and global health. *Yale J. Health Pol'y L. & Ethics*, *7*, 273.

56. Mills, A. (2014). Health care systems in low-and middle-income countries. *New England Journal of Medicine*, *370*(6), 552-557.

57. Mitropoulos, K., Cooper, D. N., Mitropoulou, C., Agathos, S., Reichardt, J. K. V., Al-Maskari, F., … Patrinos, G. P. (2017). Genomic Medicine Without Borders: Which Strategies Should Developing Countries Employ to Invest in Precision Medicine? A New "Fast-Second Winner" Strategy. OMICS: A Journal of Integrative Biology, 21(11), 647–657. doi:10.1089/omi.2017.0141

58. Morley, C., Hodge, L., Clarke, J., McIntyre, H., Mays, J., Briese, J., & Kostecki, T. (2024). 'THIS UNPAID PLACEMENT MAKES YOU POOR': Australian social work students' experiences of the financial burden of field education. Social Work Education, 43(4), 1039-1057.

59. Mudiyanselage, S. B., Stevens, J., Toscano, J., Kotowicz, M. A., Steinfort, C. L., Hayles, R., & Watts, J. J. (2023). Cost-effectiveness of personalised telehealth intervention for chronic disease management: A pilot randomised controlled trial. *PLoS ONE*, *18*(6), e0286533. https://doi.org/10.1371/journal.pone.0286533

60. Musanabaganwa, C., Ruton, H., Ruhangaza, D., Nsabimana, N., Kayitare, E., Muvunyi, T. Z., ... & Mutesa, L. (2023). An Assessment of the Knowledge and Perceptions of Precision Medicine (PM) in the Rwandan Healthcare Setting. Journal of Personalized Medicine, 13(12), 1707.

61. O'Shea, R., S, A., MA, Jamieson, R. V., & Rankin, N. M. (2022). Precision medicine in Australia: now is the time to get it right. *The Medical Journal of Australia*, *217*(11), 559–563. https://doi.org/10.5694/mja2.51777

62. Oduoye, M. O., Fatima, E., Muzammil, M. A., Dave, T., Irfan, H., Fariha, F. N. U., ... & Elebesunu, E. E. (2024). Impacts of the advancement in artificial intelligence on laboratory medicine in low-and middle-income countries: Challenges and recommendations—A literature review. Health Science Reports, 7(1), e1794.

63. Owolabi P, Adam Y, Adebiyi E. Personalizing medicine in Africa: current state, progress and challenges. Front Genet. 2023 Sep 19;14:1233338. doi: 10.3389/fgene.2023.1233338. PMID: 37795248; PMCID: PMC10546210.

64. Parekh, A. E., Shaikh, O. A., Simran, N., Manan, S., & Hasibuzzaman, M. A. (2023). Artificial intelligence (AI) in personalized medicine: AI-generated personalized therapy regimens based on genetic and medical history: short communication. *Annals of Medicine and Surgery*, *85*(11), 5831–5833. https://doi.org/10.1097/ms9.0000000000001320

65. Pastorino, R., De Vito, C., Migliara, G., Glocker, K., Binenbaum, I., Ricciardi, W., & Boccia, S. (2019). Benefits and challenges of Big Data in healthcare: an overview of the European initiatives. *European Journal of Public Health*, *29*(Supplement_3), 23–27. https://doi.org/10.1093/eurpub/ckz168

66. Peters, D. H., Garg, A., Bloom, G., Walker, D. G., Brieger, W. R., & Rahman, M. H. (2008). Poverty and access to health care in developing countries. *Annals of the New York Academy of Sciences*, *1136*(1), 161–171. https://doi.org/10.1196/annals.1425.011

67. Pilosof, N. P., Barrett, M., Oborn, E., Barkai, G., Pessach, I. M., & Zimlichman, E. (2021). Inpatient Telemedicine and New Models of Care during COVID-19: Hospital Design Strategies to Enhance Patient and Staff Safety. *International Journal of Environmental Research and Public Health*, *18*(16), 8391. https://doi.org/10.3390/ijerph18168391

68. Philippine Council for Health Research and Development (DOST-PCHRD). (2022, March 31). UP-PGC now ready for genome sequencing in VisMin - Philippine Council for Health Research and Development. Philippine Council for Health Research and Development. https://www.pchrd.dost.gov.ph/news_and_updates/up-pgc-now-ready-for-genome-sequencing-in-vismin/

69. Philippine Genome Center Mindanao. PGC Mindanao. (2023, January 18). Retrieved from https://pgc.upmin.edu.ph/

70. Rahman, A., Hossain, M. S., Muhammad, G., Kundu, D., Debnath, T., Rahman, M., Khan, M. S. I., Tiwari, P., & Band, S. S. (2022). Federated learning-based AI approaches in smart healthcare: concepts, taxonomies, challenges and open issues. *Cluster computing*, 1–41. Advance online publication. https://doi.org/10.1007/s10586-022-03658-4

71. Rajkumar, S. V. (2020). The high cost of prescription drugs: causes and solutions. *Blood Cancer Journal*, *10*(6). https://doi.org/10.1038/s41408-020-0338-x

72. RAWLS, J. (1971). *A Theory of Justice: Original Edition*. Harvard University Press. https://doi.org/10.2307/j.ctvjf9z6v

73. Record, J., Ziegelstein, R., Christmas, C., Rand, C., & Hanyok, L. (2021). Delivering personalized care at a distance: How telemedicine can foster getting to know the patient as a person. *Journal of Personalized Medicine*, *11*(2), 137. https://doi.org/10.3390/jpm11020137

74. Saeed, S. A., & Masters, R. M. (2021). Disparities in Health Care and the Digital Divide. *Current psychiatry reports*, *23*(9), 61. https://doi.org/10.1007/s11920-021-01274-4

75. Sarwar, E. (2023). Social and Ethical Implications of Integrating Precision Medicine into Healthcare. In: Global Perspectives on Precision Medicine. Advancing Global Bioethics, vol 19. Springer, Cham. https://doi.org/10.1007/978-3-031-28593-6_6

76. Sethi, Y., Patel, N., Kaka, N., Kaiwan, O., Kar, J., Moinuddin, A., Goel, A., Chopra, H., & Cavalu, S. (2023). Precision Medicine and the future of Cardiovascular Diseases: A Clinically Oriented Comprehensive Review. *Journal of clinical medicine*, *12*(5), 1799. https://doi.org/10.3390/jcm12051799

77. Sertkaya, A., Beleche, T., Jessup, A., & Sommers, B. D. (2024). Costs of drug development and research and development intensity in the US, 2000-2018. *JAMA Network Open*, *7*(6), e2415445. https://doi.org/10.1001/jamanetworkopen.2024.15445

78. Shadmi, E., Chen, Y., Dourado, I., Faran-Perach, I., Furler, J., Hangoma, P., ... & Willems, S. (2020). Health equity and COVID-19: global perspectives. *International journal for equity in health*, *19*, 1-16.

79. Smith, S. J., Smith, A. B., Kennett, W., & Vinod, S. K. (2022). Exploring cancer patients', caregivers', and clinicians' utilisation and experiences of telehealth services during COVID-19: A qualitative study. *Patient Education and Counseling*, *105*(10), 3134–3142. https://doi.org/10.1016/j.pec.2022.06.001

80. Spanakis, M., Patelarou, A. E., & Patelarou, E. (2020). Nursing personnel in the era of personalized healthcare in clinical practice. *Journal of personalized medicine*, *10*(3), 56.

81. Stefanicka-Wojtas, D., & Kurpas, D. (2023). Personalised Medicine—Implementation to the Healthcare system in Europe (Focus Group discussions). *Journal of Personalized Medicine*, *13*(3), 380. https://doi.org/10.3390/jpm13030380

82. Sun, W., Lee, J., Zhang, S., Benyshek, C., Dokmeci, M. R., & Khademhosseini, A. (2018). Engineering Precision Medicine. *Advanced science (Weinheim, Baden-Wurttemberg, Germany)*, *6*(1), 1801039. https://doi.org/10.1002/advs.201801039

83. Taruscio, D., Salvatore, M., Lumaka, A., Carta, C., Cellai, L. L., Ferrari, G., ... & Posada, M. (2023). Undiagnosed diseases: needs and opportunities in 20 countries participating in the undiagnosed diseases network international. *Frontiers in Public Health*, *11*, 1079601.

84. Tatar, M., Shoorekchali, J. M., Faraji, M. R., Seyyedkolaee, M. A., Pagán, J. A., & Wilson, F. A. (2022). COVID-19 vaccine inequality: A global perspective. *Journal of global health*, *12*, 03072. https://doi.org/10.7189/jogh.12.03072

85. Tawfik, S. M., Elhosseiny, A. A., Galal, A. A., William, M. B., Qansuwa, E., Elbaz, R. M., & Salama, M. (2023). Health inequity in genomic personalized medicine in underrepresented populations: a look at the current evidence. *Functional & Integrative Genomics*, *23*(1). https://doi.org/10.1007/s10142-023-00979-4

86. Thamlikitkul, L., Parinyanitikul, N., & Sriuranpong, V. (2024). Genomic medicine and cancer clinical trial in Thailand. Cancer Biology & Medicine, 21(1), 10.

87. Thomas, V. M., Linden, H., Gralow, J., Van Loon, K., Williams, L., Dubard-Gault, M., & Lopez, A. M. (2023). Precision medicine in Low- and Middle-Income countries. *International Journal of Cancer Care and Delivery*, *3*(2). https://doi.org/10.53876/001c.88504

88. Toyin-Thomas, P., Ikhurionan, P., Omoyibo, E. E., Iwegim, C., Ukueku, A. O., Okpere, J., ... & Wariri, O. (2023). Drivers of health workers' migration, intention to migrate and non-migration from low/middle-income countries, 1970–2022: a systematic review. BMJ global health, 8(5), e012338

89. Trabelsi, N., Othman, H., Bedhioufi, H., Chouk, H., El Mabrouk, H., Mahdouani, M., ... & H'mida, D. (2024). Is Tunisia ready for precision medicine? Challenges of medical genomics within a LMIC healthcare system. *Journal of Community Genetics*, *15*(4), 339-350

90. Trein, P., & Wagner, J. (2021). Governing personalized health: A scoping review. *Frontiers in genetics*, *12*, 650504.

91. Vellekoop, H., Versteegh, M., Huygens, S., Ramos, I. C., Szilberhorn, L., Zelei, T., Nagy, B., Tsiachristas, A., Koleva-Kolarova, R., Wordsworth, S., & Mölken, M. R. (2022). The Net Benefit of Personalized Medicine: A Systematic Literature review and regression analysis. *Value in Health*, *25*(8), 1428–1438. https://doi.org/10.1016/j.jval.2022.01.006

92. Venne, J., Busshoff, U., Poschadel, S., Menschel, R., Evangelatos, N., Vysyaraju, K., & Brand, A. (2020). International consortium for personalized medicine: an international survey about the future of personalized medicine. *Personalized medicine*, *17*(2), 89–100. https://doi.org/10.2217/pme-2019-0093

93. Vicente, A.M., Ballensiefen, W. & Jönsson, JI. How personalised medicine will transform healthcare by 2030: the ICPerMed vision. *J Transl Med* 18, 180 (2020). https://doi.org/10.1186/s12967-020-02316-w

94. Walters, S., Aldous, C., & Malherbe, H. (2023). Healthcare practitioners' knowledge, attitudes and practices of genetics and genetic testing in low- or middle-income countries - A scoping review. *Research Square (Research Square)*. https://doi.org/10.21203/rs.3.rs-2077021/v1

95. Wong, C. H., Li, D., Wang, N., Gruber, J., Lo, A. W., & Conti, R. M. (2023). The estimated annual financial impact of gene therapy in the United States. *Gene Therapy*, *30*(10–11), 761–773. https://doi.org/10.1038/s41434-023-00419-9

96. World Bank Country and Lending Groups. (n.d.). *World Bank Country and Lending Groups – World Bank Data Help Desk*. https://datahelpdesk.worldbank.org/knowledgebase/articles/906519-world-bank-country-and-lending-groups

97. Workman, P., Draetta, G. F., Schellens, J. H., & Bernards, R. (2017). How much longer will we put up with $100,000 cancer drugs? *Cell*, *168*(4), 579–583. https://doi.org/10.1016/j.cell.2017.01.034

98. World Health Organization. Global Strategy on Human Resources for Health: Workforce 2030. Strategy Report. 2016.

99. Wouters, O. J., McKee, M., & Luyten, J. (2020). Estimated research and development investment needed to bring a new medicine to market, 2009-2018. *JAMA*, *323*(9), 844. https://doi.org/10.1001/jama.2020.1166

100. Yenet, A., Nibret, G., & Tegegne, B. A. (2023). Challenges to the Availability and Affordability of Essential Medicines in African Countries: A Scoping Review. ClinicoEconomics and Outcomes Research, 15, 443–458. https://doi.org/10.2147/CEOR.S413546

101. Zerhouni, E. (2019). GAVI, the Vaccine Alliance. *Cell*, *179*(1), 13–17. https://doi.org/10.1016/j.cell.2019.08.026

102. Zhang, F., Kreuter, D., Chen, Y., Dittmer, S., Tull, S., Shadbahr, T., ... & Schönlieb, C. B. (2024). Recent methodological advances in federated learning for healthcare. *Patterns*, *5*(6).

103. Zhou, Y., Peng, S., Wang, H., Cai, X., & Wang, Q. (2024). Review of Personalized Medicine and pharmacogenomics of Anti-Cancer Compounds and natural products. *Genes*, *15*(4), 468. https://doi.org/10.3390/genes15040468